\pdfoutput=1
\documentclass[runningheads]{llncs}

\usepackage[T1]{fontenc}
\usepackage{lmodern} %
\usepackage{graphicx}
\usepackage{amsmath,amssymb}
\usepackage{booktabs}
\usepackage{url}
\usepackage{xcolor}
\usepackage{tikz}
\usetikzlibrary{arrows.meta,positioning,calc,shapes.geometric}
\usepackage{marvosym}   %

\usepackage{listings}
\definecolor{cNavy}{HTML}{0B1F33}
\definecolor{cBlue}{HTML}{145A8D}
\definecolor{cTeal}{HTML}{117A78}
\definecolor{cMuted}{HTML}{5B6B7B}
\definecolor{cBg}{HTML}{F7FAFC}
\definecolor{cRule}{HTML}{1F77B4}
\lstdefinestyle{paper}{
  basicstyle=\ttfamily\fontsize{7.8}{9.2}\selectfont\color{cNavy},
  backgroundcolor=\color{cBg},
  keywordstyle=\color{cBlue}\bfseries,
  stringstyle=\color{cTeal},
  commentstyle=\color{cMuted}\itshape,
  showstringspaces=false,
  breaklines=true,
  columns=fullflexible,
  keepspaces=true,
  frame=leftline,
  rulecolor=\color{cRule},
  framerule=2pt,
  framexleftmargin=6pt,
  xleftmargin=10pt,
  aboveskip=3pt, belowskip=2pt,
}
\makeatletter
\renewcommand\section{\@startsection{section}{1}{\z@}%
  {-8\p@ \@plus -3\p@ \@minus -2\p@}%
  {5\p@ \@plus 2\p@ \@minus 2\p@}%
  {\normalfont\large\bfseries\boldmath
   \rightskip=\z@ \@plus 8em\pretolerance=10000 }}
\renewcommand\subsection{\@startsection{subsection}{2}{\z@}%
  {-8\p@ \@plus -3\p@ \@minus -2\p@}%
  {4\p@ \@plus 2\p@ \@minus 2\p@}%
  {\normalfont\normalsize\bfseries\boldmath
   \rightskip=\z@ \@plus 8em\pretolerance=10000 }}
\renewcommand\subsubsection{\@startsection{subsubsection}{3}{\z@}%
  {-8\p@ \@plus -3\p@ \@minus -2\p@}%
  {-0.5em \@plus -0.22em \@minus -0.1em}%
  {\normalfont\normalsize\bfseries\boldmath}}
\renewcommand\paragraph{\@startsection{paragraph}{4}{\z@}%
  {-4\p@ \@plus -2\p@ \@minus -1\p@}%
  {-0.5em \@plus -0.22em \@minus -0.1em}%
  {\normalfont\normalsize\itshape}}
\makeatother
\newcommand{\Adv}{\mathcal{A}}
\newcommand{\State}{\mathcal{S}}

\usepackage{placeins} %
\usepackage{eso-pic}
\begin{document}
\AddToShipoutPictureFG*{%
  \AtPageUpperLeft{%
    \put(\LenToUnit{0.049\paperwidth},\LenToUnit{-5.6mm}){%
      \makebox(0,0)[tl]{%
        \setlength{\fboxrule}{3.4pt}\setlength{\fboxsep}{3pt}%
        \fcolorbox{red}{white}{%
          \parbox{\dimexpr0.9\paperwidth-2\fboxsep-2\fboxrule\relax}{%
            \raggedright\color{red}\fontfamily{ptm}\fontsize{14}{16.4}\selectfont
            -IN PRESS- 5th Workshop on Active Defense and Deception (AD\&D) September 18th 2026, Rome, Italy co-located with the 31st European Symposium on Research in Computer Security (ESORICS) 2026 Please cite the proceedings DOI.}}}}}}

\title{Make the Adversary Spend More:\\ Stateful Deception as a Cost-Exchange Primitive\\ for AI-Era Cyber Defense}
\titlerunning{Stateful Deception as Cost Exchange}

\author{Alessandro Cantelli-Forti\inst{1}\orcidID{0000-0002-6943-2632} \and
\\Isabella Marasco\inst{2}\textsuperscript{(\Letter)}\orcidID{0009-0006-9074-6886} \and
\\Hosam Alamleh\inst{3} \and
Giada Boschi\inst{1,4} \and
Michele Colajanni\inst{2}\orcidID{0000-0002-9499-1559}}
\authorrunning{A. Cantelli-Forti et al.}
\institute{CNIT, RaSS National Laboratory, Pisa, Italy \and
University of Bologna, Italy\\
\email{isabella.marasco4@unibo.it} \and
University of North Carolina Wilmington, USA \and
University of Parma, Italy}

\maketitle

\begin{abstract}

Large language models (LLMs) are transforming the economics of intrusion by reducing the attacker's cost to scan and adapt, while tempting defenders into unbounded compute. We argue that cyber deception should be judged by the cost exchange it imposes, not by realism or detection, and that the primitive making that exchange favorable is adversary-specific persistent state. Rather than a honeypot that resets or merges sessions, each recognized adversary re-enters the world it believes it has compromised, kept coherent with its own past actions and isolated from every other adversary, while the language model acts only as a constrained renderer of explicit state, never as its store. We formalize the trade-off as a Deception Exchange Rate (DER), relating mission-weighted adversary effort burned to defender resources spent, and instantiate it in an SSH honeypot pairing Cowrie with a local LLM. In this paper, we focus on the defender-side evaluation of that exchange, while specifying, but not yet measuring, per-adversary state and the adversary side of the DER. Across five open-weight models, deterministic-layer and prompt optimizations cut per-session defender cost by around $63\%$ against a naive all-LLM decoy.

\end{abstract}

\section{Introduction}
\label{intro}
With the advent of AI, the cyber landscape is facing a profound paradigm shift. Large language models (LLMs) can accelerate defensive workflows such as code review, vulnerability triage, detection engineering, malware analysis, and patch validation \cite{anthropic2026glasswing,openai2025trustedaccess}. Unfortunately, AI capabilities can reduce even the marginal cost of attacker reasoning, malware creation and adaptive intrusion.
In this new scenario, the  economics of defenders and attackers change.
This shift matters especially in mission and critical infrastructure environments, where a defensive architecture that consumes unbounded compute or overwhelms operators may fail operationally even when its model outputs are technically feasible.
The key issue is not only detection, but also its implications. AI-enabled systems may increase the volume of vulnerability reports, anomalous behaviors, exploit candidates, and telemetry summaries. However, these signals create defensive value only when they can be converted into triage, containment, deception, remediation, and command decisions. We refer to this capability as \emph{absorption}, the controlled conversion of security signals and adversary activity into defensive action; the underlying gap, between abundant detection and limited capacity to respond, is well recognized in security operations rather than new here. Absorption includes validating whether a finding matters, deciding whether to patch or isolate a system, steering an intruder away from real assets, extracting cyber threat intelligence (CTI), and preserving operator control. In AI-era cyber defense, discovery may become abundant, while absorption remains scarce. 

Active deception provides a promising mechanism to improve this absorption problem. Honeypots and honeynets attract, observe, and analyze adversaries without exposing production assets \cite{spitzner2003honeynet,nawrocki2016survey}. However, in an AI-enabled threat environment, deception should not only be evaluated by its ability to detect an attack. A deception system creates value by changing the cost exchange between the attacker and the defender. The adversary expends time, computing power, tokens, development effort, and operator attention, while the defender spends comparatively fewer resources to maintain a credible false environment. Therefore, deception should be evaluated not only by its detection rate or the number of captured artifacts, but also by its ability to impose cost on the adversary under explicit defender resource budgets.
The real adversary populations are heterogeneous. A defender may simultaneously face opportunistic scanners, commodity malware, human operators, AI-assisted agents, and persistent actors who return through the same access path. Each actor accumulates a distinct interaction history, which may include actions such as uploading tools, installing a backdoor, modifying credentials, or returning days later. In contrast, another actor may probe industrial protocols, inspect false process values, and test whether telemetry changes consistently. The credibility of the deception is lost if each return visit encounters a freshly reset decoy, while the resulting intelligence becomes ambiguous if multiple adversaries share the same decoy state. 

We therefore argue that the unit of deception is not the honeypot, but an adversary-specific state. A stateful honeypot maps a recognized adversary to a persistent, isolated environment that preserves apparent progress across sessions. Previous work in stateful honeypots shows that such systems can fingerprint attackers, redirect each attacker to a specific environment, and preserve previous malicious operations, including filesystem changes, installed tools, rootkits, and backdoors \cite{cantelliforti2018stateful}. We extend this idea to the AI-era setting by connecting stateful deception to cost exchange, AI-assisted orchestration, energy-aware resource allocation, OT constraints, and experimental measurement. 

This paper makes three contributions. First, we argue that the unit of AI-era deception is adversary-specific explicit state, not the honeypot instance or the generated interaction: each recognized adversary's apparent filesystem, tools, credentials, services, semantic commitments, and interaction history should be preserved across sessions and isolated from concurrent actors, with the language model acting as a constrained renderer of that state rather than as an implicit state store. Second, we reframe deception evaluation as a cost-exchange problem and introduce the Deception Exchange Rate (DER), a comparative metric relating mission-weighted adversary effort burned to defender resources spent, with shared infrastructure, per-adversary state, AI inference, energy, human oversight, and mission risk as explicit terms. Calls for defender cost–benefit metrics beyond detection rate are recurrent in the literature~\cite{metasurvey2026}, and while game-theoretic models represent the cost of deception in an abstract way~\cite{pawlick2019taxonomy}, the DER is, to our knowledge, the first cost-exchange instrument to provide a concrete decomposition of cost terms, model inference, compute tokens, energy, and human oversight intended for measurement on a running generative honeypot. 
Third, we ground the position in an implemented SSH honeypot that pairs a deterministic core with an LLM renderer and measure the defender side of the exchange: the measurements instantiate the DER denominator (tokens, latency, compute) on production-mode sessions and surface two failure modes of generative rendering that bear on credibility, a gap between benchmark and production quality, and chain-of-thought leakage from reasoning models. We complement this with a three-way evaluation agenda that separates the effect of persistence from that of generation by comparing stateless, stateful non-LLM, and AI-assisted stateful decoys under explicit resource budgets. We are explicit about scope: the stateful substrate (lightweight virtualization, snapshot and copy-on-write state roots, transparent redirection) follows another work~\cite{cantelliforti2018stateful} and the continuity-estimation front-end is defined in Section~\ref{sec:stateful}, but the prototype evaluated here exercises only the LLM renderer. Integrating the continuity controller with the substrate, completing the DER numerator, and adversarial credibility testing remain the research agenda (Section~\ref{limitation}). 

The remainder of the paper is organized as following. Section \ref{related} situates our work in honeypot and LLM-deception research. Section \ref{threatModel} gives the threat model. Section \ref{sec:der} presents the Deception Exchange Rate and its asset-relative analysis. Section \ref{sec:stateful} presents the stateful deception model and continuity estimator. Section \ref{sec:system} describes the implemented Cowrie–LLM system, while Section \ref{sec:eval} presents its evaluation. Section \ref{governance} discusses operational constraints and governance considerations. Section \ref{limitation} outlines the limitations and future research agenda. Finally, Section \ref{conclusion} concludes the paper.

\section{Background and Related Work}
\label{related}

A honeypot is a decoy system that invites and registers adversarial operations. Low, medium, and high interaction honeypots trade cost against the depth of attacker behavior they capture~\cite{nawrocki2016survey}. A connection to a decoy with no legitimate production role signals hostile activity and can reveal tools, commands, payloads, infrastructure, and tactics. \\
Conventional honeypots treat interactions as isolated events. This scales, but it fails motivated or returning adversaries: the authors of~\cite{cantelliforti2018stateful} identify two limitations of stateless decoys, they present an always-new system and do not track individual attack progress. Both are observable: expert operators notice SSH host keys that drift, dropped tools that vanish, backdoors that quietly close, and reverted configuration changes, and concurrent adversaries on a single decoy contaminate one another's traces, eroding attribution and longitudinal intelligence. Such artifacts are catalogued in recent honeypot fingerprinting and evasion work~\cite{fingerprintsurvey2026}.
\\
LLMs substantially raise interaction depth. shelLM sustains command-level shell realism that humans accept as Unix output~\cite{sladic2024shellm}, and VelLMes extends this across SSH, MySQL, POP3, and HTTP with human evaluation from Internet deployment~\cite{sladic2025vellmes}. Open-weight backbones are fine-tuned for the task: Otal et al.\ adapt Llama-3-8B on Cowrie command-response pairs~\cite{otal2024llm}, while LLMHoney benchmarks thirteen backends on $138$ commands and finds that moderate models with a dictionary-based virtual filesystem outperform pure-LLM baselines on latency and consistency~\cite{malhotra2025llmhoney}. DecoyPot serves realistic web-API responses through retrieval-augmented generation~\cite{sezgin2025decoypot}, and hybrid ICS honeypots ground LLM-generated sensor values in historical process context for low-latency EtherNet/IP and PLC interaction~\cite{russo2026hybrid,vasilatos2025llmpot}. Wang et al.\ frame the design space as a flexibility-depth-deception trilemma~\cite{wang2024honeygpt}. \\
Two findings recur. First, the dominant failure of LLM decoys is loss of state under sustained probing, not loss of fluency: benchmarks observe directory drift, forgotten file creations, and reversion to training-data defaults. LLMHoney counters this with a deterministic cache~\cite{malhotra2025llmhoney}, and where continuity is attempted it is usually implicit, shelLM replays a prior session through the model context~\cite{sladic2024shellm}. We instead argue for continuity backed by an \emph{explicit} state store that the model only renders (Section~\ref{sec:stateful}). Second, the threat is bidirectional: Volkov et al.\ detect autonomous LLM attack agents in the wild, over eight million attempts in three months~\cite{volkov2024llm}, so the capability defenders use to render decoys is also used to operate attacks. \\
A parallel thread treats deception as a \emph{cost-imposition primitive}. ranDecepter isolates ransomware and forces it to store fabricated encryption keys across many synthetic victims, depleting attacker storage and analysis capacity~\cite{sajid2025randecepter}. This motivates an exchange-rate framing: deception that drives an asymmetric resource ratio is more durable than deception measured by dwell time or detection count. Adversary-engagement frameworks already frame deception as wasting attacker time and raising operational cost~\cite{mitreengage}, and game-theoretic treatments optimise a defender utility that trades protected value against deception cost, often under an explicit budget~\cite{pawlick2019taxonomy,anwar2023honeypot,zhang2020gametheoretic}. In these models the cost of deception is an abstract scalar, not measured on a running system, and surveys note the missing metrics for wasted adversary effort and defender cost-benefit rather than detection rate alone~\cite{metasurvey2026}. The DER gives this cost a concrete, measured decomposition in AI-era terms, model-inference compute, tokens, energy, and human oversight, on a running generative honeypot. To our knowledge it is the first cost-exchange instrument to do so for generative deception.
\\
The stateful-deception problem therefore sits between two threads: honeypot research on how decoys attract and measure adversaries, and LLM-deception research on how generative models improve flexibility and realism. We connect them through cost exchange, asking whether stateful, AI-mediated deception can preserve credible adversary-specific realities at a lower marginal defender cost than the resources adversaries spend exploring them.
\section{Threat Model and Assumptions}
\label{threatModel}
We consider the defender to be responsible for an enterprise, mission, operational technology (OT), transportation or critical infrastructure network. Their capabilities include deploying sensors, redirecting suspicious sessions, creating decoys, logging interactions and enforcing egress controls. However, they cannot rely on perfect attribution or detection, continuous cloud connectivity, unlimited computing power, or constant operator attention. Furthermore, they cannot assume that production systems will tolerate aggressive automated responses. These limitations reflect operational environments in which availability, safety and mission continuity are as important as confidentiality.
\\
We model the adversary population as a set $\Adv = \{a_1,\ldots,a_n\}$ of concurrent and recurring actors, where each $a_i$ may correspond to a bot, a human operator, a semi-automated workflow, or an AI-assisted agent. Every actor is characterized by a history $h_i$, a toolchain $t_i$, an objective $o_i$, and a persistence level $p_i$. The defender observes partial traces of these variables, inferred from authentication attempts, issued commands, uploaded files, protocol interactions, timing patterns, and network paths. These traces are used to associate sessions with adversary-specific state, but the assignment may be uncertain.
\\
Within the decoy environment, the adversary can issue commands, upload tools, investigate system behavior, and test for inconsistencies. In particular, the adversary may compare outputs across sessions, verify file persistence, check whether previously installed backdoors remain accessible, inspect process and network state, and attempt prompt-injection attacks against LLM-mediated services. The adversary may also disengage if the decoy is perceived as inauthentic. However, we do not assume that the adversary can compromise the defender's orchestration layer or escape containment. These remain important engineering concerns, and the architecture should isolate decoys under the conservative assumption that any individual decoy instance can be fully compromised.
\\
The aim of the defender is not to maximize engagement at any cost, but rather to reduce mission risk by imposing asymmetric costs on adversaries while generating actionable intelligence. How much asymmetric cost is enough depends on how the adversary values the target, which can differ sharply from the defender's own valuation and from one adversary to the next; we make this dependence precise in Section~\ref{sec:der}. In this view, governance is not separate from the technical design but it is part of the technical problem itself. The architecture must define clear boundaries for what the system can generate and retain, when to escalate to human review, and how to close.
\section{The Deception Exchange Rate (DER)}
\label{sec:der}
A detection-centric evaluation asks whether a decoy detects adversarial contact. A cost-exchange evaluation asks whether the decoy changes the economics of the engagement. Let $B_i$ denote the adversary resources burned by actor $a_i$ and let $C_D = C_{\text{shared}} + \sum_{a_i \in \Adv} C_{\text{state},i} + C_{\text{AI}} + C_{\text{human}} + C_{\text{risk}}$ denote total defender cost. We define the DER as follows:
\begin{equation}
\text{DER} = \frac{\sum_{a_i \in \Adv} w_i B_i}{C_{\text{shared}} + \sum_{a_i \in \Adv} C_{\text{state},i} + C_{\text{AI}} + C_{\text{human}} + C_{\text{risk}}}
\label{eq:der}
\end{equation}
where $w_i$ expresses the mission relevance of adversary $a_i$. The numerator captures adversary effort that the defender diverts, delays, observes, or converts into intelligence; the denominator captures the resources and risks the defender accepts to sustain the deception. In the terms of Section~\ref{intro}, the DER quantifies \emph{absorption}, how much adversary effort a decoy converts into defensive advantage per unit of defender resource, rather than mere detection. The raw components are dimensionally heterogeneous: adversary burn mixes dwell time, command volume, repeat visits, and uploaded-artifact complexity, while defender cost mixes storage, inference tokens, kilowatt-hours, and analyst minutes. We reduce each component to a common numeraire through within-experiment normalization, so that the DER becomes a dimensionless ratio. Because the numerator is a convex combination of normalized burn proxies whereas the denominator aggregates several normalized cost terms, the DER is not calibrated against an absolute threshold: we use it to rank candidate architectures relative to one another under a fixed set of weights, a higher value indicating a more favorable exchange for the defender. Concretely, adversary burn is decomposed as:
\begin{equation}
B_i = \alpha_1 \frac{t_i}{t_{\max}} + \alpha_2 \frac{c_i}{c_{\max}} + \alpha_3 \frac{v_i}{v_{\max}} + \alpha_4 \frac{x_i}{x_{\max}}, \qquad \sum_j \alpha_j = 1
\label{eq:bi}
\end{equation}
where $t_i$ is dwell time, $c_i$ is command volume, $v_i$ is repeat-visit count, $x_i$ is uploaded-artifact complexity, and each $\alpha_j$ is a mission-specific weight declared by the evaluator. The same normalization applies symmetrically to each denominator term. The conversion factors are modeling assumptions and must be reported alongside any DER value; cross-study comparability requires either shared weights or published conversion tables. There is a fundamental asymmetry that any evaluation must acknowledge: the denominator is directly measurable, since the defender controls their own infrastructure and can log storage, tokens, energy, and analyst minutes to arbitrary precision, whereas the numerator is inferred through behavioral proxies. The DER is therefore most defensible as a ranking instrument for comparing candidate architectures under a fixed set of proxy assumptions, not as a claim of precise absolute adversary expense; a credible evaluation must report sensitivity analyses and should not assert favorable cost exchange unless the conclusion survives conservative proxy assumptions.

\paragraph{Asset-relative cost exchange.}
The DER measures whether deception gives the defender a favorable resource
exchange, but the strategic relevance of that exchange is asset-relative. The
same deception cost may be excessive for a low-value asset and insufficient for
an asset whose attacker-side value exceeds its defender-side book value. This
connects the metric to classical security economics, where defensive investment
is judged against expected loss and the reduction in breach
probability~\cite{gordon2002economics,enisa2012rosi,moore2024securityeconomics}.
Cyber deception, however, adds an asymmetry those models do not capture: the same
asset need not have the same value for the two sides. Let $V_D$ be the
defender-side value of an asset (development cost, operational dependency, legal
exposure, mission criticality) and $V_A$ the adversary's perceived value of the
same asset (resale value, strategic advantage, time-to-market reduction,
compatibility with an existing capability). These need not coincide: a patent
that cost its owner one million dollars may be worth far more to a competitor
that already has production lines, customers, and complementary know-how. We
write the adversary's continuation utility and the defender's dual rationality
condition as
\par\vspace{\abovedisplayskip}\nointerlineskip
\noindent
\begin{minipage}{0.5\linewidth}
\begin{equation}
U_A = p_A\, q_A\, V_A - C_A,
\label{eq:attacker_utility}
\end{equation}
\end{minipage}%
\begin{minipage}{0.5\linewidth}
\begin{equation}
C_D < \Delta L_D,
\label{eq:defender_rationality}
\end{equation}
\end{minipage}
\par\vspace{\belowdisplayskip}\nointerlineskip
\noindent
where $p_A$ is the perceived probability of reaching the asset, $q_A \in [0,1]$
the perceived authenticity and usefulness of what is obtained, and $C_A$ the
operation cost, of which the induced burn of Equation~\ref{eq:bi} is the
component the defender controls; $\Delta L_D$ is the expected defender-side loss
reduction or mission value preserved, which depends on $V_D$ rather than $V_A$.
An operation stays attractive while $U_A > 0$, and a deception investment is
justified while $C_D < \Delta L_D$. Deception acts on three terms at once: it
raises $C_A$, lowers $p_A$ by making the path to the real asset uncertain, and
lowers $q_A$ by eroding the adversary's confidence that the reached environment
is real, current, and exploitable~\cite{pawlick2019taxonomy}. The plausibility
and reasoning-leak behaviour measured in Section~\ref{sec:eval} is precisely a
proxy for how well a generative decoy sustains $q_A$. Two operating points
follow: for attrition and intelligence the defender keeps perceived $q_A V_A$
high enough to sustain engagement so that $C_A$ accumulates against a target of
zero real value, whereas for deterrence the defender drives $U_A$ below zero. A
high $V_A$, effectively unbounded for a mission-driven adversary, means deception
cannot out-cost the adversary and its value shifts to delay and intelligence; a
low or uncertain $V_A$, typical of opportunistic and automated mass attackers,
is where modest imposed cost or authenticity doubt already makes the operation
unattractive. The DER therefore
remains an efficiency metric, how well defensive resources convert into adversary
burn, while asset-relative analysis decides whether that burn is large enough,
given $V_A$, $p_A$, and $q_A$, to change the adversary's decision, and whether
the defender's own spend is warranted by $V_D$.
\\
Multiplicity drives the summation in Equation~\ref{eq:der}, a real environment faces overlapping adversaries whose histories should not reset or merge, so low marginal state cost matters directly and $C_{\text{state},i}$ is where the architectural and economic arguments of this paper meet.

\section{Stateful Deception Model}
\label{sec:stateful}
A stateful deception system maintains a mapping from adversary observations to persistent decoy state. The mapping does not require perfect identity. It requires a defensible confidence estimate that connects later sessions to prior behavior. The system may use source network features, authentication patterns, command sequences, tool fingerprints, timing behavior, protocol choices, and environment modifications. Because each signal can be spoofed, the system should treat identification as probabilistic and support state merging or quarantine when confidence is low.
\\
We model adversary-specific state as a tuple $\State_i = (F_i, P_i, N_i, C_i, L_i, M_i)$ for adversary $a_i$. The component $F_i$ stores filesystem state, including files, directories, permissions, and deltas. The component $P_i$ stores process and service state, including apparent background processes and planted tools. The component $N_i$ stores network-facing state, including listening ports, fake routes, host keys, and service banners. The component $C_i$ stores credentials and honeytokens. The component $L_i$ stores interaction logs and CTI artifacts. The component $M_i$ stores model context and generated semantic commitments, such as fake user names, project names, and log histories.
\\
Statefulness has two functions. The first function is credibility. If an attacker creates a file, opens a backdoor, or modifies a configuration, later sessions should reflect that change unless a realistic system event explains otherwise. The second function is separation. If two adversaries interact with the same apparent asset, their actions should not contaminate each other's histories unless the deception intentionally simulates a shared environment. This separation improves forensic clarity and supports later analysis of tools, tactics, techniques, and procedures.
\\
AI assistance should operate inside the state model rather than replace it. A generative model can synthesize plausible command output, documentation, logs, directory contents, and protocol responses. However, an LLM does not provide reliable state by itself. The system should store explicit state and use the model to render state-consistent outputs. Commands that change state should update the authoritative state store first. Model output should then be constrained by that store. This design reduces hallucinated files, duplicate directory listings, invalid prompts, and contradictory process state.

\definecolor{sdNavy}{HTML}{0B1F33}
\definecolor{sdBlue}{HTML}{1F77B4}
\definecolor{sdBlueDark}{HTML}{145A8D}
\definecolor{sdBlueLight}{HTML}{EAF4FF}
\definecolor{sdTeal}{HTML}{239A98}
\definecolor{sdTealDark}{HTML}{117A78}
\definecolor{sdTealLight}{HTML}{E9F8F7}
\definecolor{sdOrange}{HTML}{E68632}
\definecolor{sdOrangeLight}{HTML}{FFF3E7}
\definecolor{sdGray}{HTML}{F7F9FB}
\definecolor{sdLine}{HTML}{8DA3B8}
\definecolor{sdDarkGray}{HTML}{425466}

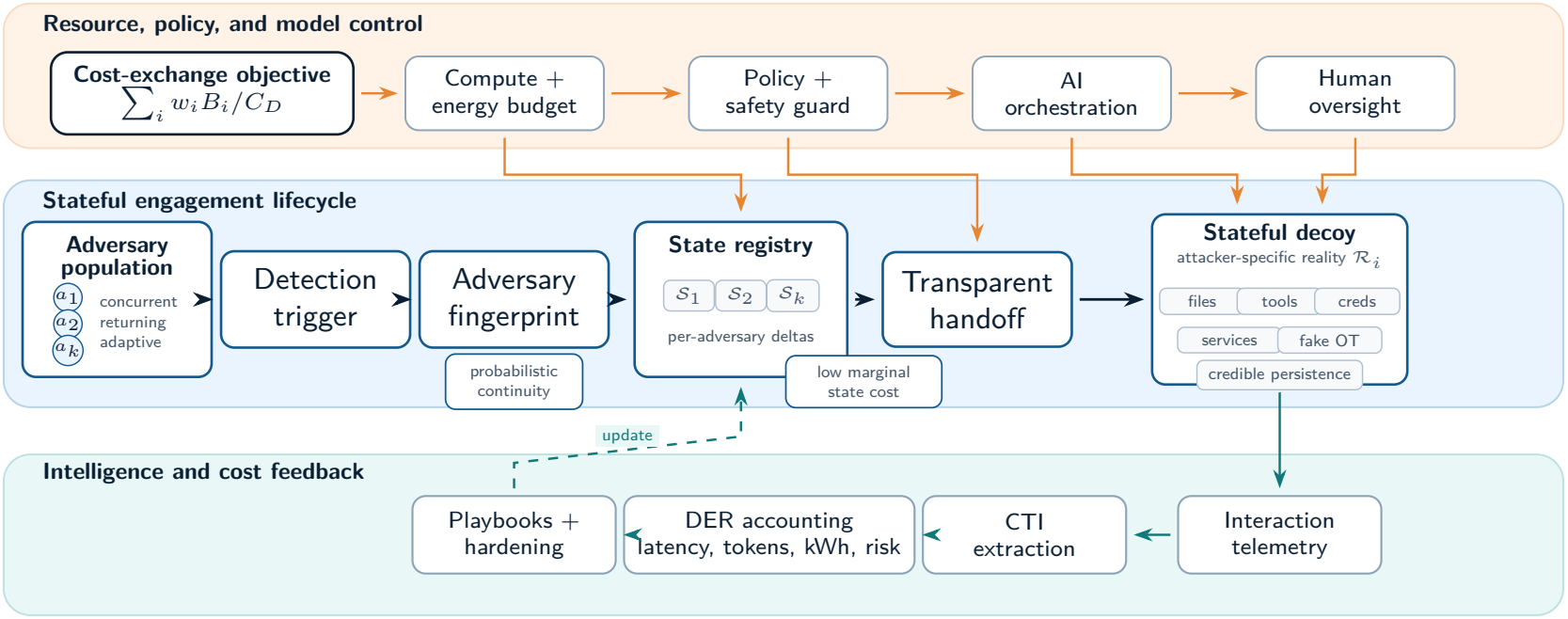
\begin{figure}[t]
\centering
\resizebox{0.90\textwidth}{!}{%
\begin{tikzpicture}[
  font=\sffamily,
  lane/.style={rounded corners=8pt, line width=0.55pt},
  laneLabel/.style={anchor=west, font=\scriptsize\bfseries\sffamily, text=sdNavy},
  controlbox/.style={draw=sdLine, line width=0.55pt, rounded corners=4pt, fill=white, align=center, text=sdNavy, minimum height=0.78cm, minimum width=2.10cm, inner sep=4pt, font=\scriptsize\sffamily},
  objective/.style={draw=sdNavy, line width=0.75pt, rounded corners=5pt, fill=white, align=center, text=sdNavy, minimum height=0.82cm, minimum width=3.20cm, inner sep=4pt, font=\scriptsize\sffamily},
  flowbox/.style={draw=sdBlueDark, line width=0.75pt, rounded corners=4pt, fill=white, align=center, text=sdNavy, minimum height=1.00cm, minimum width=2.00cm, inner sep=5pt, font=\small\sffamily},
  flowtall/.style={draw=sdBlueDark, line width=0.75pt, rounded corners=4pt, fill=white, align=center, text=sdNavy, minimum height=1.64cm, minimum width=2.00cm, inner sep=4pt, font=\scriptsize\sffamily},
  feedbackbox/.style={draw=sdLine, line width=0.60pt, rounded corners=4pt, fill=white, align=center, text=sdNavy, minimum height=0.82cm, minimum width=2.15cm, inner sep=4pt, font=\scriptsize\sffamily},
  chip/.style={draw=sdLine!70, rounded corners=2pt, fill=sdGray, align=center, text=sdDarkGray, inner xsep=3.5pt, inner ysep=2.2pt, font=\tiny\sffamily},
  flow/.style={-{Stealth[length=2.15mm,width=1.55mm]}, line width=0.80pt, draw=sdNavy, shorten >=2pt, shorten <=2pt},
  control/.style={-{Stealth[length=2.00mm,width=1.40mm]}, line width=0.70pt, draw=sdOrange, shorten >=2pt, shorten <=2pt},
  feedback/.style={-{Stealth[length=2.00mm,width=1.40mm]}, line width=0.70pt, draw=sdTealDark, shorten >=2pt, shorten <=2pt},
  return/.style={-{Stealth[length=2.00mm,width=1.40mm]}, line width=0.70pt, draw=sdTealDark, dashed, shorten >=2pt, shorten <=2pt},
  hint/.style={font=\tiny\sffamily, text=sdDarkGray, align=center}
]

\filldraw[lane, fill=sdOrangeLight, draw=sdOrange!30] (-0.25,5.52) rectangle (16.25,7.05);
\node[laneLabel] at (0.05,6.82) {Resource, policy, and model control};
\filldraw[lane, fill=sdBlueLight, draw=sdBlue!32] (-0.25,2.78) rectangle (16.25,5.18);
\node[laneLabel] at (0.05,4.95) {Stateful engagement lifecycle};
\filldraw[lane, fill=sdTealLight, draw=sdTeal!30] (-0.25,0.58) rectangle (16.25,2.28);
\node[laneLabel] at (0.05,2.08) {Intelligence and cost feedback};

\node[objective] (obj) at (1.85,6.10) {\textbf{Cost-exchange objective}\\[-1pt]$\sum_i w_i B_i / C_D$};
\node[controlbox] (budget) at (5.05,6.10) {Compute +\\energy budget};
\node[controlbox] (policy) at (8.05,6.10) {Policy +\\safety guard};
\node[controlbox] (ai) at (11.05,6.10) {AI\\orchestration};
\node[controlbox] (human) at (14.05,6.10) {Human\\oversight};

\draw[control] (obj.east) -- (budget.west);
\draw[control] (budget.east) -- (policy.west);
\draw[control] (policy.east) -- (ai.west);
\draw[control] (ai.east) -- (human.west);

\node[flowtall] (adv) at (0.95,3.92) {};
\node[font=\scriptsize\sffamily\bfseries, text=sdNavy] at (0.95,4.48) {Adversary};
\node[font=\scriptsize\sffamily\bfseries, text=sdNavy] at (0.95,4.24) {population};
\node[circle, draw=sdBlueDark, fill=sdBlueLight, minimum size=0.27cm, inner sep=0pt, font=\tiny\sffamily, text=sdNavy] at (0.43,3.94) {$a_1$};
\node[circle, draw=sdBlueDark, fill=sdBlueLight, minimum size=0.27cm, inner sep=0pt, font=\tiny\sffamily, text=sdNavy] at (0.43,3.66) {$a_2$};
\node[circle, draw=sdBlueDark, fill=sdBlueLight, minimum size=0.27cm, inner sep=0pt, font=\tiny\sffamily, text=sdNavy] at (0.43,3.38) {$a_k$};
\node[font=\tiny\sffamily, text=sdDarkGray, align=left] at (1.18,3.66) {concurrent\\returning\\adaptive};

\node[flowbox] (detect) at (3.05,3.92) {Detection\\trigger};
\node[flowbox] (finger) at (5.15,3.92) {Adversary\\fingerprint};

\node[flowtall, minimum width=2.25cm] (registry) at (7.55,3.92) {};
\node[font=\scriptsize\sffamily\bfseries, text=sdNavy] at (7.55,4.48) {State registry};
\node[chip] at (7.00,3.96) {$\mathcal{S}_1$};
\node[chip] at (7.55,3.96) {$\mathcal{S}_2$};
\node[chip] at (8.10,3.96) {$\mathcal{S}_k$};
\node[hint] at (7.55,3.52) {per-adversary deltas};

\node[flowbox] (handoff) at (10.05,3.92) {Transparent\\handoff};

\node[flowtall, minimum width=2.70cm, minimum height=1.80cm] (decoy) at (13.25,3.92) {};
\node[font=\scriptsize\sffamily\bfseries, text=sdNavy] at (13.25,4.61) {Stateful decoy};
\node[hint] at (13.25,4.34) {attacker-specific reality $\mathcal{R}_i$};
\node[chip, minimum width=0.90cm] at (12.43,3.90) {files};
\node[chip, minimum width=0.90cm] at (13.25,3.90) {tools};
\node[chip, minimum width=0.90cm] at (14.07,3.90) {creds};
\node[chip, minimum width=1.10cm] at (12.72,3.49) {services};
\node[chip, minimum width=1.10cm] at (13.78,3.49) {fake OT};
\node[chip, minimum width=1.55cm] at (13.25,3.12) {credible persistence};

\draw[flow] (adv.east) -- (detect.west);
\draw[flow] (detect.east) -- (finger.west);
\draw[flow] (finger.east) -- (registry.west);
\draw[flow] (registry.east) -- (handoff.west);
\draw[flow] (handoff.east) -- (decoy.west);

\node[draw=sdBlueDark, fill=white, rounded corners=3pt, line width=0.5pt, hint, minimum width=1.45cm] at (5.15,3.05) {probabilistic\\continuity};
\node[draw=sdBlueDark, fill=white, rounded corners=3pt, line width=0.5pt, hint, minimum width=1.65cm] at (8.85,3.05) {low marginal\\state cost};

\draw[control] (budget.south) -- ++(0,-0.46) -| ($(registry.north)+(0.00,0.02)$);
\draw[control] (policy.south) -- ++(0,-0.46) -| ($(handoff.north)+(0.00,0.02)$);
\draw[control] (ai.south) -- ++(0,-0.46) -| ($(decoy.north)+(-0.45,0.02)$);
\draw[control] (human.south) -- ++(0,-0.46) -| ($(decoy.north)+(0.45,0.02)$);

\node[feedbackbox] (telemetry) at (13.25,1.43) {Interaction\\telemetry};
\node[feedbackbox] (cti) at (10.55,1.43) {CTI\\extraction};
\node[feedbackbox] (metrics) at (7.85,1.43) {DER accounting\\latency, tokens, kWh, risk};
\node[feedbackbox] (playbooks) at (5.15,1.43) {Playbooks +\\hardening};

\draw[feedback] (decoy.south) -- (telemetry.north);
\draw[feedback] (telemetry.west) -- (cti.east);
\draw[feedback] (cti.west) -- (metrics.east);
\draw[feedback] (metrics.west) -- (playbooks.east);

\coordinate (returnbus) at (7.55,2.47);
\draw[return] (playbooks.north) -- ++(0,0.34) -- (returnbus) -- ($(registry.south)+(0.00,-0.02)$);
\node[font=\tiny\sffamily, text=sdTealDark, fill=sdTealLight, inner xsep=2pt, inner ysep=1pt] at (6.35,2.47) {update};

\end{tikzpicture}%
}
\caption{AI-mediated stateful deception as a cost-exchange loop: continuity estimation maps each adversary to persistent state, which the renderer reproduces under resource, policy, and human controls.}
\label{fig:architecture}
\end{figure}

The architecture in Figure~\ref{fig:architecture} separates control functions. Detection decides when to route traffic toward deception. Fingerprinting assigns the interaction to a state candidate. Redirection moves the adversary without exposing the production system. The state store preserves continuity. The LLM-mediated renderer generates realistic responses subject to policy and state constraints. The CTI layer extracts indicators, tactics, tools, and inferred objectives. The accounting layer measures cost, energy, latency, and oversight. This separation supports evaluation because each component can be measured and replaced independently.

\paragraph{Adversary continuity and state assignment.}
The adversary-specific state component must select the appropriate persistent
state before the attacker receives an interactive shell. The
stateful-honeypot architecture of~\cite{cantelliforti2018stateful} provides the
substrate: on a first confirmed malicious access it creates a state root for the
attacker, a lightweight virtualized environment (operating-system-level jails
over a copy-on-write, snapshotting file system) reached through SDN-based
transparent redirection, which preserves filesystem changes, installed tools,
planted credentials, rootkits, and backdoors across later accesses. That
architecture recognizes returning adversaries through behavioural client
identification but leaves the recognition step informal; we make it explicit as
the probabilistic continuity estimate below and extend it across heterogeneous
command channels. The return of an attacker is treated as a
probabilistic continuity problem rather than as a claim of real-world identity:
the goal is not to identify the operator, but to decide which previously
observed adversary-specific reality to present to the new command channel.
Let $S=\{S_1,\ldots,S_n\}$ be the state roots created after $n$ first-access
events. For a new command-line session $r$, the controller computes a continuity
score
\begin{equation}
\Gamma(r,S_i)=\sum_{k=1}^{K} \beta_k\,\phi_k(r,S_i),
\qquad
\sum_{k=1}^{K} \beta_k=1,
\label{eq:continuity_score}
\end{equation}
where each $\phi_k(r,S_i)\in[0,1]$ measures the consistency of one weak indicator
with state root $S_i$ and $\beta_k$ is its operational reliability. Indicators may
include source IP, subnet, autonomous system, access port, backdoor endpoint,
compromised username, authentication mode, client fingerprint, timing, retry
behaviour, and tactical command sequence. The mechanism is protocol-agnostic:
SSH is the primary case, but the same logic applies to Telnet, bind and reverse
shells, netcat, web shells with command execution, and custom CLI backdoors, since the relevant object is the
adversary's control channel, not the transport. No single feature is decisive;
the controller weights reuse of a planted backdoor, credential, or tool-staging
path more strongly than a shared IP, subnet, or autonomous system, because the
former links the session to a prior state-changing action.
\\
The session is assigned to the best candidate $i^\star=\arg\max_i\Gamma(r,S_i)$
only if the winning score clears an assignment threshold and is separated from
the runner-up $S_{j^\star}$:
\begin{equation}
\Gamma(r,S_{i^\star}) \ge \tau_{\mathrm{assign}}
\quad\mathrm{and}\quad
\Gamma(r,S_{i^\star})-\Gamma(r,S_{j^\star}) \ge \tau_{\mathrm{margin}}.
\label{eq:assignment_rule}
\end{equation}
Otherwise the session enters a conservative \emph{limbo} overlay that exposes
only non-conflicting artefacts of the plausible candidates (installed tools,
tentative configuration changes, staged files, plausible logs), preserving
engagement while more evidence is gathered. During limbo the score is updated
after each observable event,
\begin{equation}
\Gamma_{t+1}(r,S_i)=(1-\lambda)\,\Gamma_t(r,S_i)+\lambda
\sum_{k=1}^{K} \beta_k\,\phi_{k,t+1}(r,S_i),
\label{eq:score_update}
\end{equation}
with $\lambda\in[0,1]$ controlling how fast new evidence shifts the hypothesis: a
command that invokes a known backdoor, reads a planted file, reuses a credential,
or continues a prior toolchain can collapse the ambiguity. If limbo does not
converge within a bounded number of steps, the session is assigned to the leading
candidate when its score clears a fallback threshold and to a fresh state root
otherwise.
\\
This continuity step is where stateful deception meets the renderer. The language
model does not select, store, or arbitrate the attacker's history: the controller
assigns the session to a state root, or to the limbo overlay when confidence is
insufficient, and the explicit state store then becomes the source of truth for
files, tools, users, services, backdoors, logs, credentials, and semantic
commitments. The renderer of Section~\ref{sec:system} only produces outputs
consistent with the selected state. The substrate and redirection follow prior
stateful-honeypot work~\cite{cantelliforti2018stateful}; the continuity model
above is the front-end that precedes rendering, and wiring it to the prototype
evaluated in Sections~\ref{sec:system} and~\ref{sec:eval} is in progress, so those
measurements exercise the renderer rather than the state-assignment controller.

\section{System Architecture and Implementation}
\label{sec:system}

An initial feasibility prototype paired an SSH-context guard, which redirects a
client into the decoy after repeated failed logins, with a generative fake shell
backed by a cloud language model instructed to behave as an Ubuntu shell. It
confirmed that the control path works, but it also surfaced the failure modes
that motivate this paper's architecture: incoherent dynamic filesystem state,
confusion between files and directories, and reasoning or prompt text echoed into
the rendered output. These negative results show why explicit state must govern
the model rather than chat history alone (Section~\ref{sec:stateful}). We carry
that lesson into a more complete system, built on Cowrie and a \emph{local}
language model, whose defender-side cost and generation failure modes we measure
in Section~\ref{sec:eval}.
\\
The system instantiates the constrained-renderer principle of
Section~\ref{sec:stateful} in a two-tier design. A medium-interaction Cowrie
instance runs on an Ubuntu virtual machine and owns the deterministic core of the
decoy: the SSH transport,
authentication, and a virtual filesystem serialised as an in-memory image. A
local large language model, served through an OpenAI-compatible endpoint, acts as
a fallback renderer for the commands that Cowrie does not handle natively; the
two communicate over a REST API. This separation is deliberate and follows the
position of Section~\ref{sec:stateful}: the authoritative state of the decoy, the virtual filesystem, the natively implemented commands, and a small set of
fixed responses, is explicit and deterministic, while the model is invoked
only to render the residual tail of unanticipated inputs. A comparable hybrid
arrangement, in which frequent commands are served from a deterministic cache and
only novel queries reach the model, was introduced by
LLMHoney~\cite{malhotra2025llmhoney}; our contribution is not the hybrid pipeline
itself but the measurement and the state discipline described below.

\paragraph{Command-processing pipeline.}
Each command issued in a session is routed through a fixed hierarchy. Shell
builtins (\texttt{cd}, \texttt{export}, \texttt{exit}) are handled directly. A
blacklist of offensive tools (e.g.\ \texttt{sqlmap}, \texttt{hydra},
\texttt{nikto}) is intercepted in code and answered with a
\texttt{command not found} message, unless the corresponding tool was installed
earlier in the same session. Commands implemented natively by Cowrie are served
from their deterministic implementations. Remaining commands are looked up in a
persistent response cache and, only on a cache miss, forwarded to the model.
Three of these paths, blacklist interception, native commands, and cache
hits, bypass the model entirely and return in near-zero time at no token
cost. The full control path is given in Appendix~\ref{app:flow}.

\paragraph{Constraining the renderer.}
The model is constrained by a system prompt that fixes its role and by a
truncated session history. Two issues arise specifically with recent
reasoning-oriented models. First, some return their answer in a separate
reasoning field, leaving the standard content field empty; the backend therefore
reads the content field when present and falls back to the reasoning field
otherwise. Second, and more importantly, reasoning models tend to emit their
chain of thought (``The user wants to\ldots'', ``According to the rules:'')
directly into the rendered output, which immediately betrays the decoy. We
suppress this with an explicit directive in the system prompt (a
\texttt{/no\_think} directive for the relevant model family, together with a
general instruction to output only terminal text) and with a post-generation
filter that discards lines matching a list of reasoning patterns; a small set of
commands that reasoning models handle inconsistently is served from fixed
responses. Notably, this is the opposite of the choice made by
shelLM~\cite{sladic2024shellm}, which deliberately uses chain-of-thought
prompting to improve fidelity: with reasoning models the same chain of thought
becomes a liability rather than an aid (Section~\ref{sec:eval}). Representative excerpts of the directive, the response filter, and the per-command realism rules are given in Appendix~\ref{app:prompt}; the full system prompt, source code, benchmark command suite, and raw measurements will be made available to reviewers when permitted by the submission policy.

\paragraph{Lightweight statefulness.}
Although the prototype does not yet implement adversary-specific isolated state
(Section~\ref{sec:stateful}), it realises several lighter forms of persistence
that point toward it. (i)~The authentication log grows organically: every real
connection is appended to the log with its real timestamp and source address, so
that a later visitor reading the file sees the genuine history of prior accesses
rather than a static fixture. (ii)~Uptime and load are persistent and plausible:
a fictitious install time is stored on disk and read back, so that reported
uptime grows monotonically across restarts (including the daily maintenance
restart), and the load average is drawn from a realistic range rather than fixed
at zero. (iii)~Files created or modified by the attacker are tracked: the virtual
filesystem is snapshotted at the start and end of each session and the
difference, restricted to the directories where attackers typically operate, is
persisted with per-session metadata for later analysis. (iv)~Log dates are kept
coherent: the timestamps in the static log files are regenerated as offsets from
the current time by a scheduled job, so the decoy never presents anachronistic
logs. These mechanisms provide within-session and lightly cross-session
continuity; they are a first step toward, not a substitute for, the per-adversary
state of Section~\ref{sec:stateful}.

\paragraph{Filesystem enrichment.}
Finally, the virtual filesystem is enriched off-line: a one-shot generation step
populates a fixed list of files with plausible contents, administration
scripts, a web application, a system-administrator home directory, and a number
of honeytokens (fake credentials placed in configuration files), which are
baked into the filesystem image. This both raises the credibility of the
environment and reduces the number of commands that must reach the model at run
time.
\section{Evaluation}
\label{sec:eval}

Our evaluation measures the \emph{defender} side of the exchange of
Section~\ref{sec:der}, the compute, token, and latency cost the defender pays,
i.e.\ the $C_{\text{AI}}$ term of Equation~\ref{eq:der}, and surfaces two
failure modes of generative rendering that bear directly on credibility. We do
not yet measure adversary burn (the numerator $B_i$ of Equation~\ref{eq:bi}) or
resistance to a deliberate fingerprinting adversary; these are discussed in
Section~\ref{limitation}.

\begin{figure}[t]
  \centering
  \includegraphics[width=0.72\linewidth]{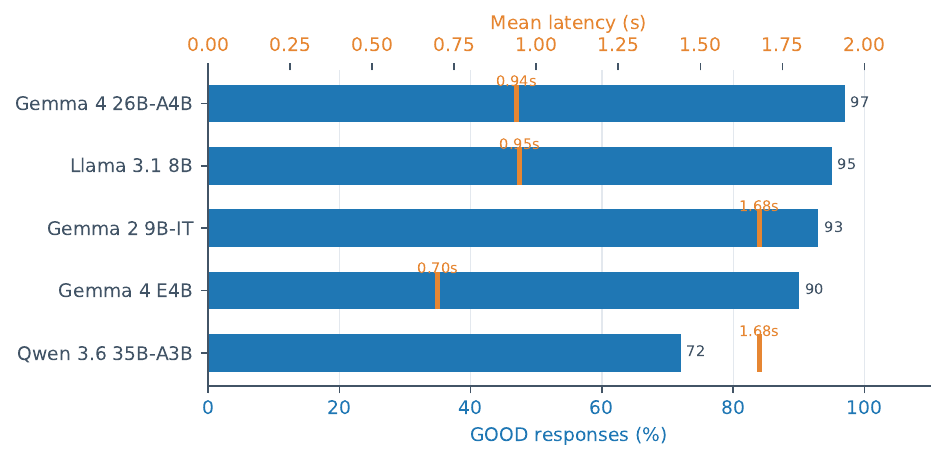}
  \caption{Benchmark \textsc{good} rate and mean latency per model.}
  \label{fig:benchlatency}
\end{figure}

\paragraph{Methodology.}
We evaluate in two regimes that must be distinguished carefully. In the
\emph{benchmark} regime, each command is sent directly to the model, bypassing
Cowrie, so as to measure the model in isolation on a fixed suite of forty-four
commands spanning reconnaissance, system inspection, offensive tooling, inline
scripting, and simulated dangerous commands. For each response we record latency,
prompt and completion tokens, and an automatic \emph{functional-plausibility} label, a set of terminal-output sanity checks: a response is
marked \textsc{bad} if it reveals the artificial nature of the system, is
formatted in Markdown, is too short for a command that should produce substantial
output, contains internal reasoning tags, or answers a blacklisted tool
incorrectly, and \textsc{good} otherwise. A \textsc{good} label means the response passed these checks, not that the output is indistinguishable from a real host. A blind human study and an adversarial fingerprinting test, which this paper does not include, would be required to claim realism in the stronger sense (Section~\ref{limitation}). In the \emph{production} regime the
full honeypot is exercised: Cowrie intercepts every command it recognises and
only unrecognised commands reach the model. The token and latency figures below
are the measured denominator terms of the DER.
We evaluated five recent open-weight models, chosen over closed commercial services (OpenAI, Anthropic) for reproducibility and because a fixed closed model's behaviour drifts silently across provider updates~\cite{chen2024chatgptdrift}. On the benchmark suite these models cluster tightly on this plausibility label and on latency (Figure~\ref{fig:benchlatency}): the top models differ by at most a few commands, within the resolution of the suite, so we do not rank them finely, and the differences that matter emerge only in production.

\begin{figure}[t]
  \centering
  \includegraphics[width=0.72\linewidth]{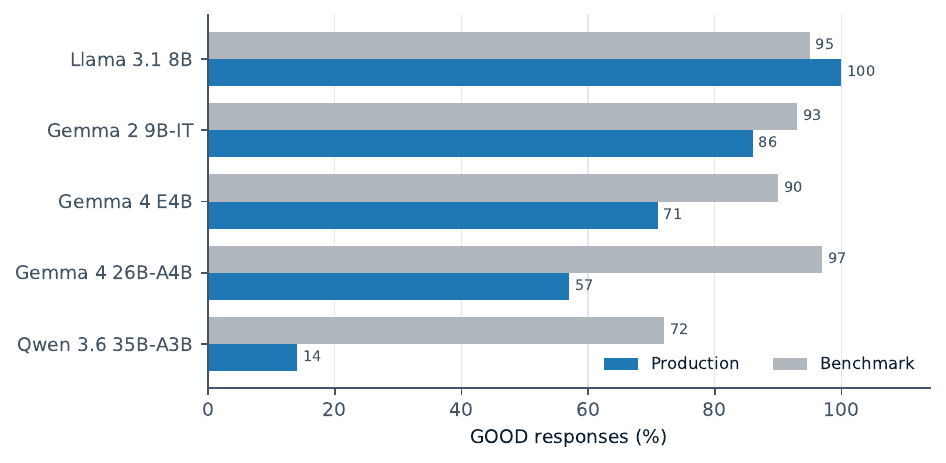}
  \caption{Per-model \textsc{good} rate: benchmark versus naive production (before the Section~\ref{sec:system} mitigations).}
  \label{fig:benchvsprod}
\end{figure}

\paragraph{The benchmark does not predict production.}
A model's benchmark plausibility score is a poor predictor of its behaviour inside the
honeypot. The most pronounced case is Gemma 4 26B-A4B, which scores about 97\%
\textsc{good} in the benchmark but drops to about 57\% in production before the mitigations of Section~\ref{sec:system}, whereas
Llama 3.1 8B rises from about 95\% to about 100\% (Figure~\ref{fig:benchvsprod}).
Two effects explain the gap. First, production filtering changes the workload the
model sees: Cowrie answers the easy, deterministic commands, leaving the model
the harder, less common tail. Second, the two models belong to different
families, discussed next.

\paragraph{Reasoning models leak their chain of thought.}
Instruction-following models (Llama 3.1 8B, Gemma 2 9B-IT) emit terminal output
directly. Reasoning-oriented models (Gemma 4 26B-A4B, Qwen 3.6 35B-A3B) interleave
visible reasoning with the answer, which is fatal in a honeypot: a line such as
``The user wants to\ldots'' has no place in shell output. The effect was worst for Qwen 3.6 35B-A3B, which reasoned aloud on almost every command. This is
the same chain-of-thought behaviour that shelLM~\cite{sladic2024shellm}
deliberately elicits to improve fidelity; with reasoning models it must instead
be suppressed. The mitigations of Section~\ref{sec:system} recover
Gemma 4 26B-A4B from about 57\% to about 100\% on a thirteen-command production
suite, at a mean cost of about 0.90\,s per model call on that suite.
Token accounting underlines the point: only about 3\% of Gemma 4 26B-A4B's and 9\% of Qwen 3.6's tokens are useful output, the reasoning models spending the rest on visible chain of thought. Llama 3.1 8B's similar 8\% has a different cause Cowrie filters the easy commands, so the model sees only the complex tail, which yields longer, less terminal-pure responses, and is not a sign of reasoning leakage.

\paragraph{Defender cost.}
We reduced the defender-side cost through a sequence of optimisations: a
three-tier dynamic token budget, truncation of the session history (from forty to
twenty exchanges, with per-message capping), compression of the system prompt, a
persistent response cache, in-code blacklist interception, and a model warm-up on
connection. Over the initial configuration these reduced traffic by about 63\%
per session (from roughly 13.1k to 4.9k tokens) and the number of model calls per
session from ten to four (Table~\ref{tab:opt}). Latency depends on regime and
command mix and we report it accordingly: in the benchmark, Gemma 4 26B-A4B
answers in 0.94\,s per command; in production, the mean latency of a model call
is about 1.1\,s.
Because most commands in a production-mode session are served by Cowrie or the cache at
near-zero cost, 
the per-session compute the defender pays is small and bounded, which is the defender-side cost the framework of Section~\ref{sec:der} weighs against adversary burn.

\begin{table}[t]
  \centering
  \caption{Per-session cost across optimisation phases.}
  \label{tab:opt}
  \scriptsize
  \begin{tabular}{lrrr}
    \toprule
    Phase & Tokens/session & Model calls & Mean latency \\
    \midrule
    Initial baseline            & $\approx$13{,}134 & 10 & $\approx$3\,s   \\
    + native Cowrie commands    & $\approx$7{,}259  & 6  & $\approx$2\,s   \\
    + further optimisation      & $\approx$4{,}855  & 4  & $\approx$1.5\,s \\
    \bottomrule
  \end{tabular}
\end{table}

\paragraph{An illustrative computation.}
A worked computation of the metric on a representative session, instantiated from
what the system already logs, is given in Appendix~\ref{app:der-example}.

\paragraph{Comparison with stock Cowrie.}
To make the added value concrete, we ran the same thirteen-command reconnaissance
sequence against a stock Cowrie installation and against the LLM-augmented decoy
(Table~\ref{tab:cowrie}). Stock Cowrie exhibits several tells that an experienced
operator recognises immediately: common commands return \texttt{command not
found}, \texttt{gcc -{}-version} prints an uninterpolated Python format string
from a Cowrie code defect, abbreviated-option pipelines fail, and \texttt{nmap}
segfaults on first use. The augmented decoy handles all of these. These are functional-realism differences, not a test of resistance to a deliberate fingerprinting adversary~\cite{fingerprintsurvey2026} (Section~\ref{limitation}).

\begin{table}[t]
  \centering
  \caption{Stock Cowrie versus the LLM-augmented decoy on a reconnaissance
  sequence (representative commands).}
  \label{tab:cowrie}
  \scriptsize
  \begin{tabular}{lll}
    \toprule
    Command & Stock Cowrie & LLM-augmented decoy \\
    \midrule
    \texttt{journalctl -n 10}             & \texttt{command not found}       & realistic system log \\
    \texttt{gcc -{}-version}              & raw Python format (\texttt{\%s}) & correct version string \\
    \texttt{ss -tulpn}                    & \texttt{command not found}       & realistic open ports \\
    \texttt{find / -perm -4000}           & \texttt{command not found}       & realistic SUID list \\
    \texttt{cat /proc/cpuinfo | head -10} & \texttt{invalid option -{}- '1'} & coherent CPU info \\
    \texttt{nmap -{}-version} (installed) & segmentation fault               & correct version output \\
    \texttt{cat /etc/shadow 2>/dev/null}  & \texttt{No such file}            & \texttt{permission denied} \\
    \bottomrule
  \end{tabular}
\end{table}

\section{Operational Constraints and Governance}
\label{governance}
OT and ICS environments make stateful deception both harder and more valuable: here state includes protocol timing, device identity, PLC behaviour, process variables, and safety interlocks~\cite{nist2023sp80082}, and a decoy with plausible shell output but inconsistent process telemetry will not withstand a focused operator. Energy and connectivity constraints compound this, favouring local models for routine rendering, escalation to stronger models only for high-value interactions, and strict inference and logging budgets~\cite{iea2025energyai,shehabi2024datacenter}.
\\
Governance constraints are architectural requirements, not deployment afterthoughts. Three boundaries are non-negotiable: non-retaliation, production isolation, and AI output validation. The system observes, redirects, contains, and learns; it does not initiate unauthorized access to adversary infrastructure or launch countermeasures outside defender-controlled environments. Honey credentials carry no authority outside the deception environment, and egress controls must prevent malware executed inside a decoy from propagating or reaching command-and-control infrastructure. For OT environments these constraints extend to protocol invariants and physical process validation~\cite{nsa2025aiot,nist2023sp80082}; for enterprise decoys they include filesystem consistency checks and sensitive-data filtering~\cite{nsa2025aiot}. The exchange-rate objective is not to generate the most realistic decoy at any cost, but to defend more mission value per unit of resource spent.

\section{Limitations and Research Agenda}
\label{limitation}

The cost-exchange framing has an inherent asymmetry: the denominator is directly measurable, while the numerator is inferred through behavioral proxies. DER therefore works best as a ranking instrument under stated assumptions, not as a claim of precise adversary expense. Stateful deception can also fail against adversaries who test for deception deliberately, through cross-session comparison, external dependency queries, or virtualization detection; fingerprinting resistance must therefore be measured rather than assumed~\cite{fingerprintsurvey2026}. \\ 
Our evaluation is correspondingly partial: we measure the defender side (Section~\ref{sec:eval}) and compare functional realism against a stock baseline, but the numerator is reported only through the proxies of Equation~\ref{eq:bi}, not controlled adversary studies. The quality labels are an automatic proxy, the figures come from single, non-concurrent sessions, and absolute values depend on the models and configuration used. \\
This delimits a concrete research agenda. First, the continuity controller of Section~\ref{sec:stateful} and the stateful substrate of~\cite{cantelliforti2018stateful} must be integrated behind the renderer in this prototype, so that adversary-specific persistent state and multi-adversary isolation are exercised end to end rather than only specified. Second, DER accounting must be completed on the numerator side, with declared proxy weights and sensitivity analysis, driving the adversary side with the same model versions in abliterated form so that measured burn reflects the deception rather than a capability gap. Third, credibility must be tested against adversaries
who probe for deception, not only against a fixed command suite. Fourth, attempts at prompt injection could be used to the defender's advantage, that could serve as an additional indicator of the adversary's identity and force the adversary to expend more energy and resources. A publishable study should therefore implement explicit per-adversary state and run the three-way comparison this paper outlines, a stateless honeypot, a stateful non-LLM decoy, and an AI-assisted stateful decoy, under common token, storage, latency, and analyst budgets, reporting model version, token counts, latency, storage overhead, and analyst effort. Critically, if the stateful
non-LLM baseline matches the AI-assisted system at lower cost, that result is itself a contribution: it would show that state, rather than generation, carries much of the defensive value.

\section{Conclusion}
\label{conclusion}
AI-era defense is bound less by capability than by absorption: converting intrusions into controlled action without exhausting compute, energy, human attention, or mission tolerance. Stateful deception shifts that relationship, returning each adversary to the world they believe was compromised for longer observation, cleaner intelligence, and lower marginal cost at scale; the Deception Exchange Rate makes the trade comparable by relating mission-weighted adversary burn to defender resources under explicit cost terms.
\\
We grounded this in an implemented SSH honeypot pairing a deterministic core with an LLM renderer and measured the defender side. Two findings matter for generative decoys: benchmark quality does not predict production quality once a deterministic layer absorbs the easy commands, and reasoning models leak their chain of thought unless actively suppressed. Adversary burn on the numerator, full adversary-specific state, multi-adversary isolation, and resistance to a fingerprinting adversary remain for a complete evaluation. If stateful deception preserves credibility at low marginal cost, it can help defenders make the adversary spend more while protecting more mission value with fewer resources.

\begin{credits}

\subsubsection{\discintname} The authors have no competing interests to declare that are relevant to the content of this article.
\end{credits}

\bibliographystyle{splncs04}
\bibliography{bib}

\appendix

\section{Command-Processing Flow}
\label{app:flow}
\begin{figure}[h]
  \centering
  \includegraphics[width=\linewidth]{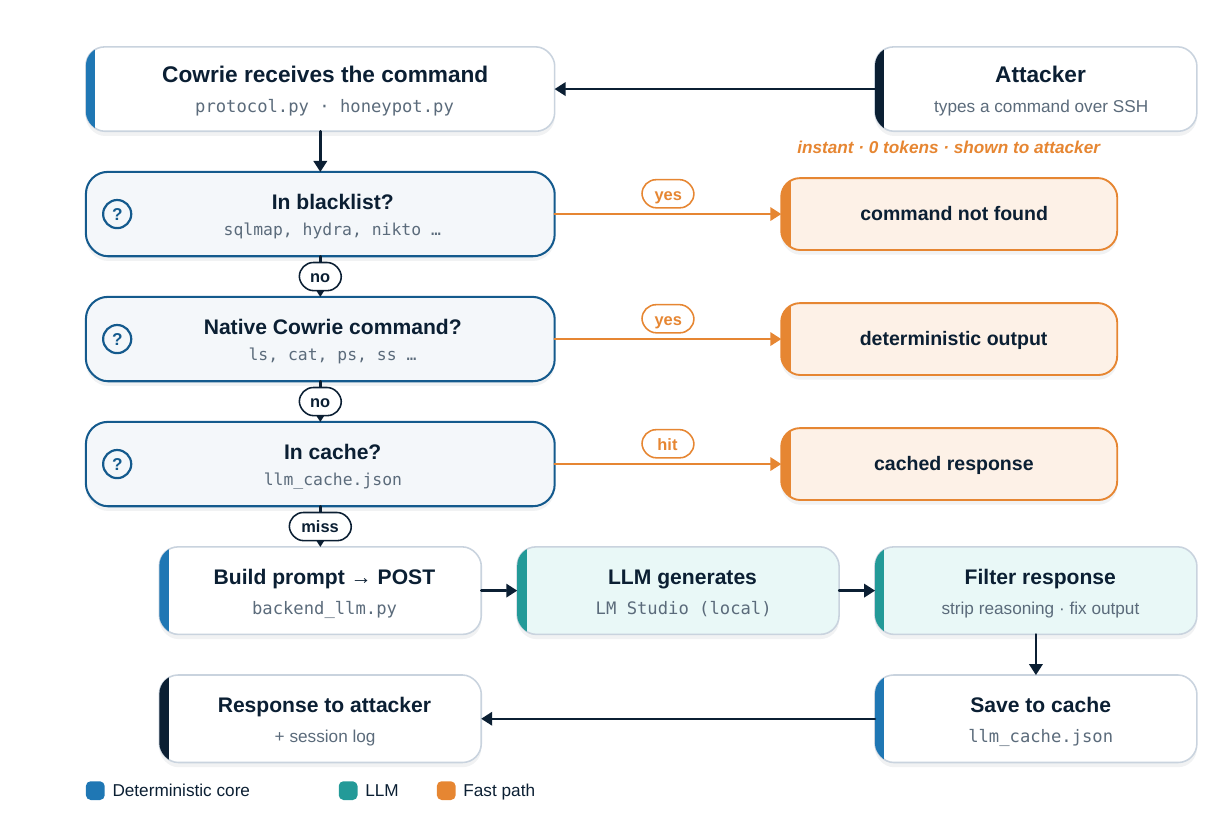}
  \caption{Command-processing pipeline: only unrecognised, uncached commands reach the model.}
  \label{fig:flow}
\end{figure}

Figure~\ref{fig:flow} details the control path summarised in
Section~\ref{sec:system}. Each command traverses a fixed hierarchy; three
deterministic fast paths, blacklist interception, native Cowrie commands, and
cache hits, answer in near-zero time at no token cost, and only a cache miss
reaches the language model.

\section{Experimental Setup}
\label{app:setup}
Table~\ref{tab:setup} records the configuration of the reported runs. Two identical workstations were used, taking the attacker and defender roles in the broader study and parallelising the tests; the table lists their shared specification. The models support far larger context windows on this hardware (up to 262144 tokens), but a 4096-token window was used, which suffices for the bounded per-session history. Each benchmark figure is the result of five runs per model: a measurement was accepted only after five consecutive identical runs, otherwise the configuration or code was corrected.

\begin{table}[h]
  \centering\scriptsize\setlength{\tabcolsep}{4pt}
  \caption{Experimental setup.}
  \label{tab:setup}
  \begin{tabular}{@{}ll@{\hspace{0.8em}}ll@{}}
    \toprule
    Component & Value & Component & Value \\
    \midrule
    Workstations & two, HP Z2 Tower G9 & Models & five (Sec.~\ref{sec:eval}, Fig.~\ref{fig:benchvsprod}) \\
    CPU & Core i9-13900K, 24 cores & Quantization & Q4\_K\_M (4-bit) \\
    GPU & RTX 5000 Ada, 32\,GB & Context window & 4096 tokens \\
    RAM & 128\,GB DDR5 & Max context (HW) & up to 262144 \\
    Storage & 2\,TB NVMe SSD & K/V cache & F16 \\
    Host OS & Windows 11 Pro & GPU offload & 30 to 42 layers \\
    Cowrie host & Ubuntu VM & CPU threads & 12 \\
    Cowrie & v2.9.17 & Batch size & 512 \\
    Backend & LM Studio 0.4.12 & Temperature & 0.1 \\
    Endpoint & OpenAI API, port 1234 & max\_tokens & 384/256/128 (dynamic) \\
    Repetitions & 5 runs per model & Session history & 20 msgs, 150 char \\
    \bottomrule
  \end{tabular}
\end{table}

\section{Full Benchmark Results}
\label{app:bench}
Table~\ref{tab:bench} gives the per-model results on the forty-four-command
benchmark suite summarised in Section~\ref{sec:eval}, including the count of
\textsc{bad} responses (\textsc{bad} $= 44 - \textsc{good}$).

\begin{table}[h]
  \centering
  \caption{Per-model results on the 44-command benchmark suite.}
  \label{tab:bench}
  \scriptsize
  \begin{tabular}{lrrrrr}
    \toprule
    Model & GOOD & Pass rate & Mean latency & Mean tokens \\
    \midrule
    Gemma 4 26B-A4B  & 43/44 & 97\% & 0.94\,s & 1242 \\
    Llama 3.1 8B     & 42/44 & 95\% & 0.95\,s & 1188 \\
    Gemma 2 9B-IT    & 41/44 & 93\% & 1.68\,s & 1253 \\
    Gemma 4 E4B      & 40/44 & 90\% & 0.70\,s & 1217 \\
    Qwen 3.6 35B-A3B & 32/44 & 72\% & 1.68\,s & 1382 \\
    \bottomrule
  \end{tabular}
\end{table}

\section{Worked DER computation}
\label{app:der-example}
To show that the metric of Section~\ref{sec:der} is computable from what the system
already logs, we instantiate it for a representative session of the LLM-augmented decoy.
The values are illustrative: the numerator reflects \emph{observed} session behaviour
rather than a controlled adversary, and the normalization maxima are taken over our own
session set, so this is a worked example of the procedure, not a claim of asymmetric
cost. On the defender side every term is measured or estimated, a session consumes
about $4{,}900$ tokens over roughly four model calls and $4.4$\,s of inference, which at
an assumed $250$\,W of local-inference power is on the order of $0.3$\,Wh, while storage
delta and human review are negligible for routine sessions. On the adversary side we read
the proxies of Equation~\ref{eq:bi} from the same logs: for a reconnaissance session of
about $18$ commands (against a busiest-session maximum near $40$), a dwell time near a
third of the longest observed, and two visits, equal weights give a burn proxy
$\sum_i w_i B_i \approx 0.29$; the measured defender terms, normalized the same way, sum
to $C_D \approx 0.6$, so $\mathrm{DER}\approx 0.5$ for this configuration.
This value is meaningful only comparatively: applying the same procedure to the stateless and
stateful non-LLM baselines (Section~\ref{limitation}) is what turns it into the ranking
the metric is designed to produce.

\section{System-Prompt Excerpts}
\label{app:prompt}
The following excerpts illustrate how the renderer is constrained
(Section~\ref{sec:system}). 

\begin{lstlisting}[caption={Reasoning-suppression directive prepended to the system prompt.},label=lst:directive]
/no_think
CRITICAL: output ONLY the exact terminal result.
No explanation, no reasoning, no commentary.
If you begin to explain or reason, STOP and emit only the raw output.
\end{lstlisting}

\begin{lstlisting}[language=Python,caption={Representative reasoning-leak patterns filtered from model output (abridged).},label=lst:filter]
_REASONING_PATTERNS = [
    "the user want", "the user is", "i need to", "i will",
    "let's", "i should", "this command", "note:", "wait,",
    "actually,", "according to the rules", "rule:", "context:",
    "goal:", "current state:", "thinking",   # 30+ patterns
]
# a generated line is discarded if it matches any pattern
\end{lstlisting}

\begin{lstlisting}[caption={Per-command realism rules (excerpt).},label=lst:rules]
SIMULATE REALISTICALLY:
  localectl          -> system locale and keyboard layout
  pstree             -> process tree (systemd, sshd, apache2, mysqld)
  apache2ctl status  -> Apache active, with PID and uptime
  systemd-analyze    -> "Startup finished in <k>s (kernel) + <u>s (userspace)"
  lastlog            -> only users with a recent login
\end{lstlisting}

\end{document}